\documentclass[preprint,amsmath,amssymb,aps]{revtex4}
\usepackage{graphicx}
\usepackage{epsfig}
\usepackage{multirow}
\begin{document}
\title{A new renormalization procedure of the quasiparticle random phase approximation}
\author{A. A. Raduta$^{1,2}$, C. M. Raduta $^{1}$}
\affiliation{$^{1}$Institute of Physics and Nuclear Engineering$\mathrm{,}$ P.O. Box MG06, Bucharest 077125, Romania\\
$^{2}$Academy of Romanian Scientists, 54 Splaiul Independentei, Bucharest 050094, Romania}

\begin{abstract}
The ground state of a many body Hamiltonian considered in the quasiparticle representation is redefined by accounting for the quasiparticle quadrupole pairing interaction. The residual interaction of the newly defined quasiparticles is treated by the QRPA. Solutions of the resulting equations exhibit specific features.  In particular, there is no interaction strength where the first root is vanishing. A comparison with other renormalization methods is presented.
\end{abstract}
\maketitle 
\renewcommand{\theequation}{1.\arabic{equation}}
\setcounter{equation}{0}
\section{Introduction}
The big merit of the liquid drop model (LDM) proposed by Bohr and Mottelson \cite{Bohr} is that one defined by the concept of rotational bands. Also, some collective properties of spherical nuclei have been nicely described. The main drawback of LDM consists of the fact that it accounts only for the spherical and harmonic motion of the drop, while many experimental data reclaims a non-harmonic picture and, moreover, many nuclei exhibit static deformed shapes.
Many phenomenological improvements have been proposed along the time, among which few are to be mentioned: a) rotation-vibration model \cite{FaGr}; b) Gneus-Greiner model \cite{GnGr}; c) generalized collective formalism \cite{Hess}
; d) coherent state model \cite{Rad1,Rad2}; e) interacting boson approximation \cite{ArIa}. In parallel, microscopic theories have been formulated trying to get counterparts of the phenomenological methods and interpret the nuclear collective motion in terms of the single particle motion. Thus, the random phase approximation built on the top of either the Hartree-Fock or the BCS ground state (QRPA) \cite{Maruhn} provides a collective state which corresponds actually to the one phonon state predicted by the harmonic LDM \cite{KissSo}. Another important result is that of Kumar and Baranger, who calculated the inertial and stiffness parameter microscopically \cite{KuBa}, the potential energy surface leading to some sound nuclear structure interpretation.
Based on the RPA ground state, several procedures of accounting for some new correlations, i.e. of going beyond RPA, have been proposed. Such procedures are related with the equations of motion method \cite{KerSha,DoDa,RaSaLi,RaSa} or boson expansion technique \cite{BeZe,MaYa,RaCeSt,KleMa}.

A distinct procedure which keeps the appealing harmonic picture of RPA but also includes in the definition of the phonon operator new correlations is obtained by renormalizing the specific equations
 \cite{Har}. Thus, the drawback of the standard RPA formalism of collapsing for a critical value of the attractive long range interaction strength is removed. Indeed, the collective root of the RPA equations goes to zero not for a finite value of the mentioned interaction strength, but only asymptotically. This approach was extended to the proton-neutron dipole interaction in Refs. 
\cite{ToiSu}. Vanishing the excitation energy of the collective RPA state corresponds to a phase transition where the ground state is unstable to adding small  contribution. Around this critical interaction the RPA method is no longer valid. In order to stabilize the ground state it is necessary either to change the mean field for the single particle motion, which results in having deformed single particle orbits, or to renormalize the basic equations.

In the present paper we propose a new method of renormalizing the QRPA equations. As we shall see, the result for the collective root is that it does not vanish in the critical  interaction strength, where the standard QRPA collapses, but reaches a minimum value and moreover  the energy increases when a subsequent increase of the strength is performed. The new point of this work is that the mean
 field is redefined in the quasiparticle picture by including in the ground state the quasiparticle quadrupole pairing correlations. As a result, both the new quasiparticles and the new QRPA solutions are deformed.

The project sketched above will be described according to the following plan. In Section 2 the model Hamiltonian is presented. For the sake of completeness the results for the standard BCS and QRPA equations are briefly described. Section 3 is devoted to the deformed quasiparticles or, in other words, to the second order BCS approach. The new BCS and QRPA equations are analytically derived.
In section 4 the formalism is numerically applied to a single $j$ shell. The final conclusions are drawn in Section 5. 
   
\renewcommand{\theequation}{2.\arabic{equation}}
\setcounter{equation}{0}
\section{The model Hamiltonian}
We consider a system of nucleons described by a many body Hamiltonian consisting of a spherical mean field term, the pairing and the quadrupole-quadrupole two body interactions. Written in second quantization, this has the form:
\begin{equation}
H=\sum_{\alpha}(\epsilon_{a}-\lambda)c^{\dagger}_{\alpha}c_{\alpha}-\frac{G}{4}P^{\dagger}P-\frac{X}{4}\sum_{\mu}Q_{2\mu}Q_{2-\mu}(-)^{\mu},
\label{modham}
\end{equation}
where $P^{\dagger}$ and $Q$ denote the pairing and quadrupole operator, respectively:
\begin{eqnarray}
P^{\dagger}&=&\sum_{\alpha}c^{\dagger}_{\alpha}c^{\dagger}_{-\alpha}(-)^{j_{\alpha}-m_{\alpha}},\nonumber\\
Q_{2\mu}&=&\sum_{\alpha ,\beta}\langle\alpha|r^2Y_{2\mu}|\beta\rangle c^{\dagger}_{\alpha}c_{\beta}\equiv \sum_{a,b,\mu}q_{ab}\left(c^{\dagger}_{a}c_{b}\right)_{2\mu},\nonumber\\
q_{ab}&=&\frac{\hat{j_a}}{\hat{2}}\langle a||r^2Y_{2\mu}||b\rangle, \;\;\rm{with} \;\;\hat{j}_a = \sqrt{2j_a+1} .
\end{eqnarray}
The particle-hole quadrupole operator is defined as:
\begin{equation}
\left(c^{\dagger}_{a}c_{b}\right)_{2\mu}=\sum_{m_{\alpha},m_{\beta}}C^{j_a j_b 2}_{m_{\alpha} -m_{\beta} \mu}c^{\dagger}_{\alpha}c_{\beta}(-)^{j_b-m_{\beta}}.
\end{equation}
The second quantization is used within the spherical shell model basis $|\alpha\rangle =|a, m_{\alpha}\rangle =|n_al_aj_am_{\alpha}\rangle$. Thus, the creation (annihilation) operator of one particle in the state $|\alpha\rangle$ is denoted by $c^{\dagger}_{\alpha} (c_{\alpha})$. We used also the notation 
$|-\alpha \rangle =|a,-m_{\alpha}\rangle $.

The sum of the first two terms is quasi-diagonalized by passing to the quasiparticle representation defined by the Bogoliubov-Valatin (BV) transformation:
\begin{eqnarray}
a^{\dagger}_{\alpha}&=&U_{\alpha}c^{\dagger}_{\alpha}-V_as_{\alpha}c_{\alpha}, \nonumber\\
a_{\alpha}&=&U_ac_{\alpha}-V_as_{\alpha}c^{\dagger}_{-\alpha},~~\rm{with}~~ s_{\alpha}=(-)^{j_a-m_{\alpha}}. 
\end{eqnarray}
The quasiparticle vacuum state will be hereafter denoted by $|BCS\rangle $.
Up to an additive constant, in the new representation the Hamiltonian is:
\begin{equation}
H=\sum_{\alpha}E_{a}a^{\dagger}_{\alpha}a_{\alpha}-\frac{X}{4}\sum_{\mu}Q_{2\mu}Q_{2,-\mu}(-)^{\mu},
\end{equation}
where $E_{a}$ denotes the quasiparticle energy for the state  characterized by the set of quantum numbers "$\alpha$", while the quadrupole operators can be expressed in terms of two quasiparticle and quasiparticle quadrupole density operators:
\begin{eqnarray}
Q_{2\mu}&=&\sum_{a\le b}q_{ab}\left[\xi_{ab}\left(A^{\dagger}_{2\mu}(ab)+A_{2-\mu}(ab)(-)^{\mu}\right)+\eta_{ab}\left(B^{\dagger}_{2\mu}(ab)+B_{2-\mu}(ab)(-)^{\mu}\right)\right],
\nonumber\\ 
A^{\dagger}_{2\mu}(ab)&=&\frac{1}{\sqrt{1+\delta_{ab}}}\sum_{m_{\alpha}, m_{\beta}}C^{j_a\;\; j_b\;\; 2}_{m_{\alpha}m_{\beta}\mu}a^{\dagger}_{\alpha}a^{\dagger}_{\beta},\nonumber\\
B^{\dagger}_{2\mu}(ab)&=&\sum_{m_{\alpha}, m_{\beta}}C^{j_a\;\; j_b\;\; 2}_{m_{\alpha}-m_{\beta}\mu}a^{\dagger}_{\alpha}a_{\beta}s_{\beta},\nonumber\\
A_{2\mu}(ab)&=&\left(A^{\dagger}_{2\mu}(ab)\right)^{\dagger};~~B_{2\mu}(ab)=\left(B^{\dagger}_{2\mu}(ab)\right)^{\dagger},\nonumber\\
\xi_{ab}&=&\frac{1}{\sqrt{1+\delta_{ab}}}\left(U_aV_b+U_bV_a\right), \eta_{ab}=\frac{1}{1+\delta_{ab}}\left(U_aU_b-V_aV_b\right).
\end{eqnarray}
In deriving the expression of the quadrupole operator in the quasiparticle representation, some symmetry properties were used:
\begin{eqnarray}
A^{\dagger}_{2\mu}(ba)&=&(-)^{j_a-j_b}A^{\dagger}_{2\mu}(ab);~~A_{2\mu}(ba)=(-)^{j_a-j_b}A_{2\mu}(ab),\nonumber\\
B^{\dagger}_{2\mu}(ba)&=&(-)^{j_a-j_b}B_{2-\mu}(ab)(-)^{\mu};~~B_{2\mu}(ba)=(-)^{j_a-j_b}B^{\dagger}_{2-\mu}(ab)(-)^{\mu},\nonumber\\
q_{ba}&=&(-)^{j_a-j_b}q_{ab}.
\end{eqnarray}

The quasiparticle many body Hamiltonian is treated within the Random Phase Approximation (QRPA) formalism.
Thus, one defines a phonon operator
\begin{equation}
C^{\dagger}_{2\mu}=\sum \left[X(ab)A^{\dagger}_{2\mu}(ab)-Y(ab)A_{2-\mu}(ab)(-)^{\mu}\right],
\end{equation}
with the amplitudes $X(ab)$ and $Y(ab)$ determined such that the following equations are fulfilled:
\begin{equation}
\left[H,C^{\dagger}_{2\mu}\right]=\omega C^{\dagger}_{2\mu},\;\;\left[C_{2\mu},C^{\dagger}_{2\mu'}\right]=\delta_{\mu,\mu'}.
\label{qrpa}
\end{equation}
The first equation yields for the phonon amplitudes the so called QRPA equations:
\begin{eqnarray}
\left(\begin{matrix}{\cal A} & {\cal B}\cr -{\cal B}^* & -{\cal A}^*\end{matrix}\right)\left(\begin{matrix} X\cr Y\end{matrix} \right)
=\omega \left(\begin{matrix} X\cr Y\end{matrix} \right).
\label{qrpaeq}
\end{eqnarray}
This is a homogeneous system of linear equations which determines the amplitudes up to a multiplicative factor which is fixed by the second equation (\ref{qrpa}), which gives:
\begin{equation}
\sum_{ab}\left[|X(ab)|^2-|Y(ab)|^2\right]=1.
\end{equation}
The matrices involved in Eq. (\ref{qrpa}) have the expressions:
\begin{eqnarray}
{\cal A}_{ab,a'b'}&=&\left(E_a+E_b\right)\delta_{a,a'}\delta_{b,b'}-\frac{X}{2}r_{ab}r_{a'b'},\nonumber\\
{\cal B}_{ab,a'b'}&=&-\frac{X}{2}r_{ab}r_{a'b'},\;\;\rm{with}\;\;r_{ab}=q_{ab}\xi_{ab}.
\end{eqnarray}
Once the QRPA equations are solved, the phonon space is defined. Thus, the vacuum state denoted by $|RPA\rangle$ is the ground state, while the excited states are multi-phonon excitations of
$|RPA\rangle$.
Since the two body interaction is of a separable form, the QRPA equations provide a dispersion equation for the excitation energies and analytical expressions for the phonon amplitudes.
By inspection of these expressions, one finds that the first excited state corresponds to an energy smaller than the minimal two quasiparticle energy and has a collective character.
Due to the attractive nature of the two body interaction, the collective state energy is decreasing when the interaction strength X is increased. Consequently, for a critical value of X the phonon energy is vanishing and the phonon operator is undetermined. This situation defines the breaking down point of QRPA approach. To avoid this regime, the mean field for the single particle motion 
should be re-defined, which results in renormalizing the ground state energy such that the collective state has a positive energy. Concretely, the spherical shell model single particle basis is to be replaced with the Nilsson single particle basis. Another way to remove the QRPA breaking down is to define a new  phonon operator by going beyond the QRPA approach, which is actually based on the quasi-boson approximation of the two quasiparticle quadrupole operators $A^{\dagger}_{2\mu}$ and $A_{2\mu}$. The quasi-boson commutation equations have been corrected by retaining from the exact expression not only the constant term, but also the scalar one which is considered in the average. This average is determined self-consistently with the QRPA solution and, consequently, the collective root energy goes to zero only asymptotically.
In next section we formulate a new method to renormalize the ground state energy.
\renewcommand{\theequation}{3.\arabic{equation}}
\setcounter{equation}{0}
\section{Deformed quasiparticles}
Here we study the BCS ground state excitation:
\begin{eqnarray}
|\widetilde{BCS}\rangle &=&e^T|BCS\rangle, \rm{with}\nonumber\\
T&=&z_{\alpha}a^{\dagger}_{\alpha}a^{\dagger}_{-\alpha}-z^{*}_{\alpha}a_{-\alpha}a_{\alpha}s_{\alpha}.  
\end{eqnarray}
For what follows it is useful to employ the polar representation of the parameters $z_{\alpha}$:
\begin{equation}
z_{\alpha}=\rho_{\alpha} e^{i\varphi}. 
\end{equation}
The images of the quasiparticle operators through the aforedefined transformation, are:
\begin{eqnarray}
d^{\dagger}_{\alpha}=e^Ta^{\dagger}_{\alpha}e^{-T}=a^{\dagger}_{\alpha}\cos(2\rho_{\alpha})-a_{-\alpha}s_{\alpha}\sin(2\rho_{\alpha})e^{-i\varphi_{\alpha}},
\nonumber\\
d_{\alpha}=e^Ta_{\alpha}e^{-T}=a_{\alpha}\cos(2\rho_{\alpha})-a^{\dagger}_{-\alpha}s_{\alpha}\sin(2\rho_{\alpha})e^{i\varphi_{\alpha}}.
\end{eqnarray}
With the obvious notations:
\begin{equation}
u_{\alpha}=\cos(2\rho_{\alpha});\;\; v_{\alpha}=\sin(2\rho_{\alpha})e^{-i\varphi_{\alpha}},
\end{equation}
we recognize the BV transformation for the quasiparticle operators.
The transformation parameters $u_{\alpha}$ and $v_{\alpha}$ satisfy the equation:
\begin{equation}
u^{2}_{\alpha}+|v_{\alpha}|^{2}=1,
\label{restrict}
\end{equation}
which reflects the fermionic character of the new quasiparticle operators $d^{\dagger}_{\alpha}$ and $d_{\alpha}$.
We note that the transformed state $|\widetilde{BCS}\rangle $ is vacuum state for the quasiparticle annihilation operators $d_{\alpha}$.
It is worth mentioning that due to the transformation dependence on the magnetic quantum number $m_{\alpha}$, the new quasiparticle operators are tensors of rank $j_{a}$ with indefinite projection.
In this respect, one can assert that the newly defined quasiparticles are deformed operators.

The parameters $u_{\alpha}$ and $v_{\alpha}$ may be viewed as classical coordinates depending on time. Moreover, considering the polar representation, the phase $\varphi_{\alpha}$, defining the coefficient $v_{\alpha}$,  has the meaning of a generalized linear momentum. Consequently, since we are concerned with the static properties of the new BV transformation coefficients, it is reasonable to consider vanishing phases $\varphi_{\alpha}$.

In what follows we try to determine  the parameters $u_{\alpha}$ and $v_{\alpha}$ such that the transformed state $|\widetilde{BCS}\rangle$ becomes the true ground state of the many body system under consideration, i.e. it corresponds to an energy lower than that associated with $|BCS\rangle$. The new ground state is a deformed function, which is reflected by the non-vanishing value of the expected quadrupole moment:
\begin{equation}
\langle \widetilde{BCS}|Q_{20}|\widetilde{BCS}\rangle = \sqrt{20}\sum_{a,m_\alpha >0}\frac{q_{aa}}{\hat{j_{a}}}C^{j_a 2 j_a}_{m_{\alpha} 0 m_{\alpha}}
\left(\xi_{aa}(u_{-\alpha}v_{\alpha}+u_{\alpha}v_{-\alpha})+\eta_{aa}(v^{2}_{\alpha}+v^{2}_{-\alpha})\right)\equiv q_{0}.
\end{equation}
The last part of the above relation expresses the fact that  the state $|\widetilde{BCS}\rangle$ has a definite quadrupole deformation $q_{0}$.
The average value of $H$ with the deformed state $|\widetilde{BCS}\rangle$ is:
\begin{eqnarray}
{\cal E'}&=&\langle\widetilde{BCS}|H|\widetilde{BCS}\rangle = \sum_{\alpha}E_{a}v^{2}_{\alpha}-\frac{\Delta^{2}_{20}}{4X}\\
        &-&\frac{X}{2}\sum_{a,b,m_{\alpha}}\left[q_{ab}C^{j_{a} j_{b} 2}_{m_{\alpha} -m_{\alpha} 0}\left(\xi_{ab}(u_{-\alpha}u_{b,m_{\alpha}}-v_{\alpha}v_{b,-m_{\alpha}})+
        \eta_{ab}(u_{b,m_{\alpha}}v_{\alpha}+u_{-\alpha} v_{b,-m_{\alpha}})\right)\right]^2,\nonumber
\label{erond}
\end{eqnarray} 
where we denoted:
\begin{equation}
\Delta_{20}=\frac{X}{2}\langle \widetilde{BCS}|Q_{20}|\widetilde{BCS}\rangle .
\end{equation}
Note that the last term of Eq. (\ref{erond}) is provided by the average of the quasiparticle terms of the type $d_{\alpha}d_{b,m_{\alpha}}d^{\dagger}_{b',m_{\alpha '}}d^{\dagger}_{-\alpha'}$.
However, such terms will be treated at the QRPA level, i.e. at a later stage. Due to this reason, hereafter, the mentioned term from Eq. (\ref{erond}) will be neglected.

In what follows we shall look for the stationary points of the function:
\begin{equation}
{\cal E}=\sum_{\alpha}E_av^{2}_{\alpha}-\frac{\Delta^{2}_{20}}{4X} -\sum_{\alpha}\mu_{\alpha}\left(u^{2}_{\alpha}+v^{2}_{\alpha}-1\right),
\end{equation}
with $\mu_{\alpha}$ denoting the Lagrange multiplier corresponding to the restriction (\ref{restrict}).
These are solutions of the equations obtained by vanishing the first derivatives of ${\cal E}$ with respect to the parameters $u_{\alpha}$ and $v_{\alpha}$, defining the BV quasiparticle transformation. Eliminating the Lagrange multipliers, one arrives at:
\begin{equation}
\Delta_{20}\xi_{aa}\bar{Q}_{\alpha\alpha}\left(u_{\alpha}u_{-\alpha}-v_{\alpha}v_{-\alpha}\right)-\left(E_a-2\Delta_{20}\eta_{aa}\bar{Q}_{\alpha\alpha}\right)u_{\alpha}v_{\alpha}=0,
\label{2BCS}
\end{equation}
where the following notation has been used:
\begin{equation}
\bar{Q}_{\alpha\alpha}=\frac{\hat{2}}{2\hat{j_a}}C^{j_{a} 2 j_{a}}_{m_{\alpha} 0 m_{\alpha}}q_{aa}.
\end{equation}
We remark that the above equation is invariant to the change $\alpha\to -\alpha$. This suggests that the solutions of the above equation satisfy:
\begin{equation}
u_{\alpha}=u_{-\alpha},\;\;v_{\alpha}=v_{-\alpha}.
\end{equation}
Thus, the occupation probabilities acquire the expressions:
\begin{eqnarray}
\left(\begin{matrix}v^{2}_{\alpha}\cr u^{2}_{\alpha}\end{matrix}\right)=\frac{1}{2}\left(1\mp\frac{E_a-2\Delta_{20}\eta_{aa}\bar{Q}_{\alpha\alpha}}
{\sqrt{\left(E_a-2\Delta_{20}\eta_{aa}\bar{Q}_{\alpha\alpha}\right)^2+\left(2\Delta_{20}\xi_{aa}\bar{Q}_{\alpha\alpha}\right)^2}}\right).
\label{v2u2}
\end{eqnarray}
It is interesting to note that by the quadrupole moment restriction and 
\begin{equation}
\Delta_{20}=\frac{q_{0}}{4X},
\end{equation}
Eq.(\ref{v2u2}) fully determines the parameters $v_{\alpha}$ and $u_{\alpha}$.
With the notations:
\begin{eqnarray}
\bar{E}_{\alpha}&=&E_a-2\Delta_{20}\bar{Q}_{\alpha\alpha},\;\;\delta_{\alpha}=2\Delta_{20}\bar{Q}_{\alpha\alpha},\nonumber\\ 
e_{\alpha}&=&\sqrt{\bar{E}^{2}_{\alpha}+\delta^{2}_{\alpha}},
\end{eqnarray}
one obtains a more transparent expression for the BV transformation coefficients:
\begin{eqnarray}
\left(\begin{matrix}v^{2}_{\alpha}\cr u^{2}_{\alpha}\end{matrix}\right)=\frac{1}{2}\left(1\mp\frac{\bar{E}_{\alpha}}{e_{\alpha}}\right).
\end{eqnarray}
The quantity $e_{\alpha}$ has the significance of the second order quasiparticle energy, i.e. the energy corresponding to the deformed quasiparticle state 
$|\alpha\rangle = d^{\dagger}_{\alpha}|\widetilde{BCS}\rangle $. 
Further, we shall define a phonon operator
\begin{equation}
\Gamma^{\dagger}=\sum_{ab,m_{\alpha}}\left(X^{ab}_{m_{\alpha}}d^{\dagger}_{\alpha}d^{\dagger}_{b,-m_{\alpha}}-Y^{ab}_{m_{\alpha}}d_{b,-m_{\alpha}}d_{\alpha}\right),
\end{equation}
such that it obeys the equations:
\begin{equation}
\left[H,\Gamma^{\dagger}\right]=\omega\Gamma^{\dagger},\;\;\left[\Gamma,\Gamma^{\dagger}\right]=1.
\end{equation}
In terms of the new quasiparticles, the model Hamiltonian is:
\begin{equation}
H=\sum_{\alpha}e_{\alpha}d^{\dagger}_{\alpha}d_{\alpha}-\frac{X}{4}\sum_{a,b,m_{\alpha}}Q^{ab}_{m_\alpha}\left(d^{\dagger}_{\alpha}d^{\dagger}_{b,-m_{\alpha}}+d_{b,-m_{\alpha}}d_{\alpha}\right)
\sum_{a',b',m_{\alpha'}}Q^{a'b'}_{m_\alpha'}\left(d^{\dagger}_{\alpha'}d^{\dagger}_{b',-m_{\alpha'}}+d_{b',-m_{\alpha'}}d_{\alpha'}\right),
\end{equation}
with the notation:
\begin{equation}
Q^{ab}_{m_{\alpha}}=C^{j_a \;\;j_b\;\; 2}_{m_{\alpha}-m_{\alpha} 0}q_{ab}\left[\xi(u_{\alpha}u_{b,-m_{\alpha}})+\eta_{ab}(u_{-\alpha}v_{b,-m_{\alpha}}+v_{\alpha}u_{b,m_{\alpha}})\right].
\label{qabm}
\end{equation}
The amplitudes $X^{ab}_{m_{\alpha}}$ and $Y^{ab}_{m_{\alpha}}$ are determined by the QRPA equations, which are of a similar form as those given by Eq. (\ref{qrpaeq}), 
and the normalization condition:
\begin{equation}
2\sum_{a,b,m_{\alpha}}\left[ \left(X^{ab}_{m_{\alpha}}\right)^2-\left(Y^{ab}_{m_{\alpha}}\right)^2\right]=1.
\end{equation}
The matrices involved in the QRPA equations have the expressions:
\begin{eqnarray}
{\cal A}^{ab;a'b'}_{m_{\alpha};m_{\alpha'}}&=&(e_{\alpha}+e_{b,m_{\alpha}})\delta{aa'}\delta_{bb'}\delta_{m_{\alpha}m_{\alpha'}}-XQ^{ab}_{m_{\alpha}}Q^{a'b'}_{m_{\alpha'}},\nonumber\\
{\cal B}^{ab;a'b'}_{m_{\alpha};m_{\alpha'}}&=&-XQ^{ab}_{m_{\alpha}}Q^{a'b'}_{m_{\alpha'}}.
\end{eqnarray}
Since the two body interaction involved in $H$ is separable, the compatibility condition for the QRPA equations may be brought to the form of a dispersion equation, while the phonon amplitudes are analytically expressed.
Now, it is worth noting that both the QRPA and the deformed BCS equations involve the factors $\eta_{ab}$. This reflects the fact that the terms $B^{\dagger}_{2\mu}$ and 
$B_{2-\mu}(-)^{\mu}$ of the Hamiltonian expressed in terms of spherical quasiparticles, contribute to the mentioned equations. This feature contrasts the standard QRPA equations, which ignore the scattering terms .

\renewcommand{\theequation}{4.\arabic{equation}}
\setcounter{equation}{0}
\section{The case of a single $j$ shell}
The essential features of the QRPA formalism with a multi-shell calculations can be recovered by restricting the single particle space to a single $j$. Since here we are not interested in quantitative details, but rather  in underlying the main virtues of the proposed formalism, we consider the numerical application for the single $j$ case. Thus,  one considers a system of N=10 nucleons moving in the spherical shell model state $j=i_{13/2}$  and described by the corresponding many body Hamiltonian (\ref{modham}).
We present separately the spherical and deformed QRPA results.
\subsection{Results for  QRPA built on the top of the first order BCS}
The occupation probabilities are:
\begin{equation}
V^2=\frac{N}{2\Omega},\;\;U^2=1-\frac{N}{2\Omega}.
\label{occpr}
\end{equation}
For the sake of simplifying the notation, the low indices of $U$ and $V$, specifying the chosen single $j$ are omitted. The state semi-degeneracy is denoted by $\Omega$.
Neglecting the term $G\Omega V^4$ accounting for the renormalization of the single particle energy due to the residual interaction, the BCS ground state energy is:
\begin{equation}
E=2\epsilon\Omega V^2-\frac{\Delta^2}{G}=2\epsilon\Omega V^2-G\Omega^2V^2(1-V^2).
\end{equation}
Here the Fermi level energy is set equal to zero. The condition of minimum energy leads to:
\begin{equation}
V^2=\frac{1}{2}\left(1-\frac{2\epsilon}{G\Omega}\right).
\label{2ndeqv2}
\end{equation}
This expression is consistent with Eq. (\ref{occpr}) if the single particle energy is:
\begin{equation}
\epsilon=\frac{G\Omega}{2}\left(1-\frac{N}{\Omega}\right).
\end{equation}
Equation (\ref{2ndeqv2}) gives for the quasiparticle energy:
\begin{equation}
E_{q}=\frac{G\Omega}{2}.
\end{equation}
In our application we took $G=0.4 MeV$, which results of having $E_{q}=1.4 MeV$.
The compatibility condition for the QRPA equations reads:
\begin{equation}
\omega^2=4E_{q}^2-2Xq^{2}_{jj}\xi^{2}_{jj}.
\end{equation}
The positive root of this equation is:
\begin{equation}
\omega = \left[G^2\Omega^2-\frac{8}{5}\Omega X\left(\langle j||r^2Y_2||j\rangle\right)^2\frac{N}{2\Omega}\left(1-\frac{N}{2\Omega}\right)\right]^{1/2}.
\end{equation}
The QRPA energy is plotted in Fig. 1 as function of $X$.
\begin{figure}
\includegraphics{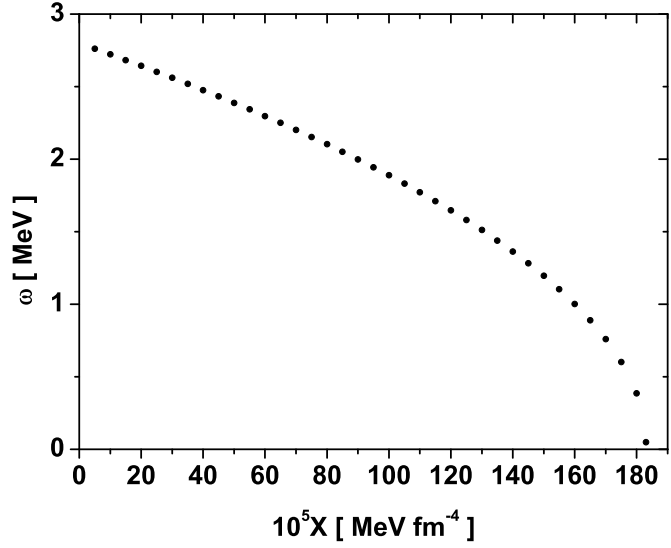}
\caption{The spherical QRPA energy as function of the quadrupole-quadrupole interaction strength for the case of a single shell, $j=i_{13/2}$.}
\end{figure}
From this figure we notice that for $X=0$ the mode  energy is equal to twice the quasiparticle energy, while for $X\approx 1.82\; 10^{-3} MeV fm^{-4}$ this is vanishing. Within this interval the function is monotonically decreasing. The vanishing mode energy reclaims a breaking down for the QRPA approach.
In next subsection we describe a method of recovering the validity of the QRPA approach.
\subsection{The study of QRPA for the quadrupole pairing correlated quasiparticles} 
Switching on the quadrupole pairing correlations for the spherical quasiparticles, we have:
\begin{eqnarray}
\bar{E}_{m}&=&E_q-\frac{X}{4}q_{0}(U^2-V^2)\langle j||r^2Y_{2}||j\rangle C^{j 2 j}_{m 0 m},\nonumber\\
\delta_{m}&=&\frac{X}{2}q_{0}\sqrt{2}UV\langle j||r^2Y_{2}||j\rangle C^{j 2 j}_{m 0 m},\\
e_{m}&=&\sqrt{\bar{E}^{2}_{m}+\delta^{2}_{m}},\nonumber\\
v^{2}_{m}&=&\frac{1}{2}\left(1-\frac{\bar{E}_{m}}{e_{m}}\right),\;\;u^{2}_{m}=1-v^{2}_{m}.
\end{eqnarray}
\begin{figure}
\includegraphics{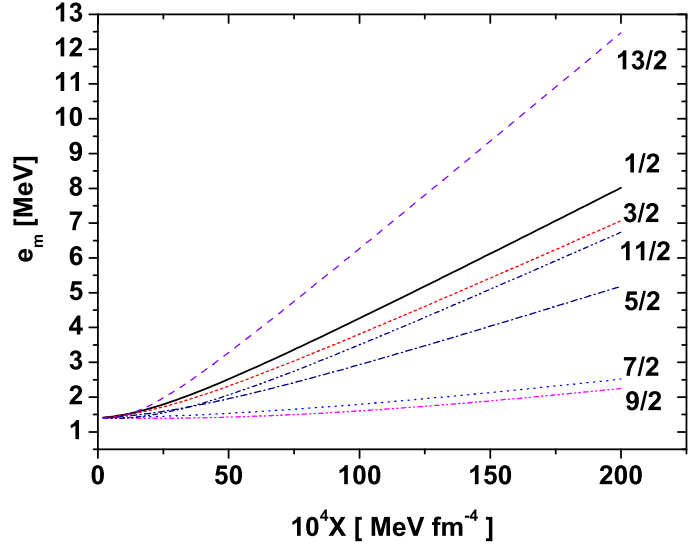}
\caption{The deformed quasiparticle energies for the $j=i_{13/2}$ multiplet.}
\end{figure}
Numerical results correspond to $q_{0}=20 fm^{2}$. One notices that the newly defined quasiparticle energy depends on the magnetic quantum number "$m$". The split, caused by the quadrupole moment of the single particle state $|jm\rangle $, is shown in Fig. 2 as function of $X$, the strength of the QQ interaction. 
Another peculiarity for the second order BCS ground state is that the average number of quasiparticles is not vanishing.
\begin{equation}
\langle\widetilde{BCS}|\hat{N}_q|\widetilde{BCS}\rangle =\sum_{m}v_{m}^{2},
\end{equation}
where the quasiparticle number operator is denoted by
\begin{equation}
\hat{N}_q=\sum_{\alpha}a^{\dagger}_{\alpha}a_{\alpha}.
\end{equation}
\begin{figure}
\includegraphics{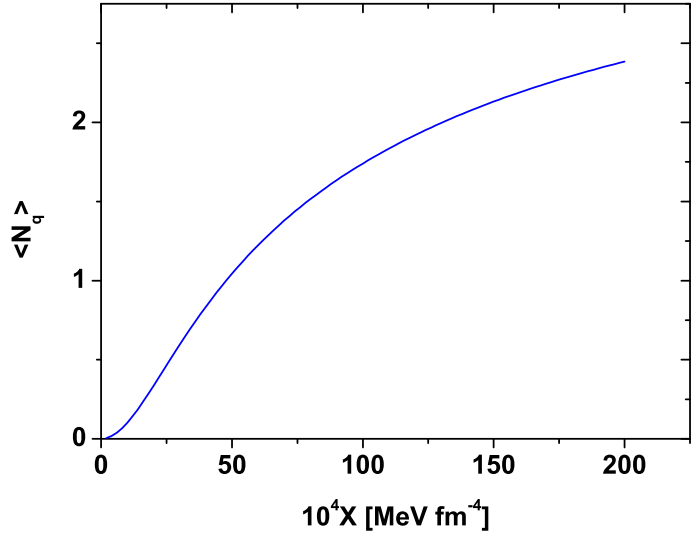}
\caption{The average number of quasiparticles  in the second order BCS state, $|\widetilde{BCS}\rangle$.}
\end{figure}
The dependence of the quasiparticle average number on the interaction strength is shown in Fig. 3. Note that the larger the strength $X$, the larger the quasiparticle averaged number.
This result implies also the presence of quasiparticles in the QRPA ground state.

In order to write the QRPA equation, we need to know the matrix
$Q^{ab}_{m_{\alpha}}$ defined by Eq. (\ref{qabm}). In the case of a single $j$ shell one obtains:
\begin{equation}
Q^{jj}_{m}=(-)^{j-m}C^{j 2 j}_{m 0 m}\langle j||r^2 Y_{2}|| j \rangle \left(\sqrt{2}UV(u_{m}^{2}-v^{2}_{m})+(U^2-V^2)u_{m}v_{m}\right).
\end{equation}

\begin{figure}
\includegraphics{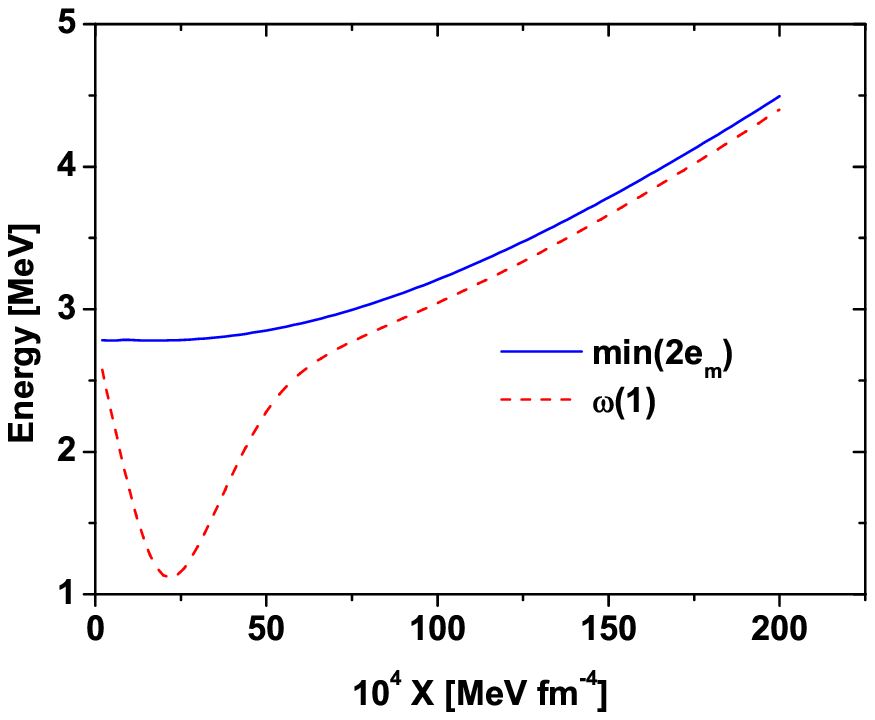}
\caption{The first QRPA equation root as a function of the QQ interaction strength. The minimal two quasiparticle energies are also presented as a function of $X$.}
\end{figure}

The compatibility condition for the QRPA equations can be written under the form of a dispersion equation:
\begin{equation}
1=X\sum_{m>0}\frac{\left(Q^{jj}_{m}\right)^{2}4e_m}{4e_{m}^{2}-\omega^{2}}.
\end{equation}
For the chosen value of $j$ there are seven solutions for $\omega$ denoted by $\omega(m)$ and ordered as:
\begin{equation}
\omega(1)<\omega(2)<...<\omega(7).
\end{equation}
The first root has a collective character since several quasiparticle pairs contribute to the phonon operator. Note that $\omega(1)$ is in magnitude smaller than the minimal two quasiparticle energy, $min(2e_{m})$. The two aforecompared quantities are represented as function of the long range interaction strength $X$, in Fig.4.
We note that $\omega(1)$ is no longer a monotonic function of $X$. There are, however, two intervals of different monotony. On the first interval $\omega(1)$ is  decreasing, reaches a minimum and then it increases in the second interval. The minimum value is reached for the value of X where the spherical phonon energy is vanishing. Due to this behavior, in the second interval there is no breaking down for the QRPA approach. Due to the specific dependence of the terms $2e_{m}$ and $Q^{jj}_{m}$ on the strength $X$, $\omega(1)$ behaves as if from the minimum point on, the effective two body interaction changes its attractive nature to a repulsive one. Also, it is worth noting that in the region around the minimum $X$ the difference $min(2e_m)-\omega(1)$
is large comparing it with the values corresponding to the $X$ from the complementary interval. This feature reflects the collective property \cite{IuRa} of the corresponding phonon state. We may say that the maximal collectivity is reached for the critical value of X. Around this point the spherical and deformed systems might be described in an unified fashion by using a spherical single particle basis.
\renewcommand{\theequation}{5.\arabic{equation}}
\setcounter{equation}{0}
\section{Summary and Conclusions}
In the previous sections we formulated an approach of renormalizing the QRPA  such that no breaking down shows up. Indeed, the first QRPA energy, instead of vanishing, it becomes minimum and then, by increasing the long range interaction strength, is increasing. Things happen as if the effective interaction changes its character, from attractive to an repulsive one. The formalism redefines first the system ground state by accounting for the quasiparticle quadrupole pairing interaction. Moreover, on the top of the newly defined ground state a QRPA  description is constructed.
It turns out that the drawback of the standard QRPA of collapsing for a critical value of the interaction strength, is removed. In the new picture some higher QRPA dynamics is included. Indeed,
the scattering terms are effectively participating in building up the new phonon operator.

We note that the new quasiparticles are not tensors of definite rank and projection. They have however a definite $j$. This makes the difference with the picture where first one defines a deformed mean field and then the pairing correlations are considered. In this case $j$ is not a good quantum number, but $\Omega$ is. This difference favors the present approach, when the QRPA is supplemented by an angular momentum  projection operation of the many body states.

The states considered in the present work are characterized by $K=0$ and, therefore, by the total angular projection, finite bands of $K=0$ can be defined.

{\bf Acknowledgment.} This work was supported by the Romanian Ministry for Education Research Youth and Sport through the CNCSIS project ID-2/5.10.2011.

\end{document}